\documentclass{Interspeech2024}

\interspeechcameraready

\title{Human Feedback Driven Dynamic Speech Emotion Recognition}

\name[affiliation={1}]{Ilya}{Fedorov}
\name[affiliation={2}]{Dmitry}{Korobchenko}

\address{
  $^1$NVIDIA, Switzerland\\
  $^2$NVIDIA, UK}
\email{ilyaf@nvidia.com, dkorobchenko@nvidia.com}


\makeatletter
\newcommand{\E}{\operatorname*{\mathbb{E}}\ilimits@}
\makeatother
\usepackage{amsmath}
\usepackage{adjustbox}
\DeclareMathOperator*{\argmax}{\arg\!\max}
\DeclareMathOperator*{\argmin}{\arg\!\min}
\begin{document}

\maketitle

\begin{abstract}
This work proposes to explore a new area of dynamic speech emotion recognition. Unlike traditional methods, we assume that each audio track is associated with a sequence of emotions active at different moments in time. The study particularly focuses on the animation of emotional 3D avatars. We propose a multi-stage method that includes the training of a classical speech emotion recognition model, synthetic generation of emotional sequences, and further model improvement based on human feedback. Additionally, we introduce a novel approach to modeling emotional mixtures based on the Dirichlet distribution. The models are evaluated based on ground-truth emotions extracted from a dataset of 3D facial animations. We compare our models against the sliding window approach. Our experimental results show the effectiveness of Dirichlet-based approach in modeling emotional mixtures. Incorporating human feedback further improves the model quality while providing a simplified annotation procedure. 
\end{abstract}

\section{Introduction}

Speech Emotion Recognition (SER) plays an important role in bridging human-computer interaction by analyzing vocal expressions to determine emotional states. Traditionally, SER assigns static emotional labels to speech recordings, facilitating applications for a wide range of sectors, from customer service to healthcare and education. This static approach, while useful, often fails to capture the fluid nature of emotional expressions over time, limiting its applicability in scenarios requiring detailed emotional understanding.

One such application emerging from recent advancements in Large Language Models (LLMs) is the creation of intelligent conversational agents. These agents, when embodied as 3D digital humans, can serve as advanced non-player characters (NPCs) in video games, providing dynamic and responsive interaction. Additionally, they can function as virtual assistants in educational or therapeutic settings, offering personalized support. This technological leap not only enriches user experience across various domains but also sets a new standard for the integration of emotional intelligence in AI systems.

We explore the realism of such digital humans with a pre-trained NVIDIA Audio2Face network \cite{audio2face} designed for animating 3D characters. This neural network generates facial animations, including movements of the lips, eyebrows, eyes, and cheeks, by processing two key inputs: the audio of human speech and its corresponding sequence of emotional states. Building upon this setup, we introduce Dynamic Speech Emotion Recognition (DSER), a novel approach that predicts sequences of emotions over time from audio. DSER extends beyond traditional SER methods, offering a new dimension of emotional intelligence to the animation of 3D characters, achieving a level of realism previously unattainable.

The development and training of DSER models face significant challenges due to the complex data labeling procedure. For a classic supervised learning approach, manually assigning emotional labels to numerous time stamps for each training audio is required. This is a meticulous process, requiring the annotator to carefully label emotions at each moment in time. Additionally, the ambiguity of emotions further complicates this process.

Given these data labeling challenges, we opted to fine-tune our models using human feedback. This approach has significantly advanced the field of LLMs, offering a substantial advantage in situations where ground truth labels are ambiguous, making it easier to identify the desired behavior rather than demonstrate it explicitly.

In this study, we propose a multi-step pipeline for training a Dynamic Speech Emotion Recognition system. The process includes classic SER model training, synthetic sequence-to-sequence data generation for DSER based on the sliding window approach, and fine-tuning of the model with human feedback using the Direct Preference Optimization algorithm \cite{DPO}. Additionally, we introduce a novel approach to formalizing the SER problem statement through Dirichlet distribution modeling. Our models are evaluated on a small dataset of emotional sequences extracted via an optimization process from a facial animation dataset. We compare our method with the sliding window approach. The results show that the proposed methods significantly improve the quality of the prediction compared to heuristics.

\section{Related works}

Our exploration into Dynamic Speech Emotion Recognition (DSER) leverages the Wav2Vec2.0 architecture \cite{Wav2Vec2}, demonstrating significant advancements over traditional neural network models like CNNs and LSTMs in predicting emotions \cite{pepino21_interspeech}. Transformer-based architectures, including Wav2Vec2.0 and HuBERT \cite{Hubert}, have been previously explored for emotion recognition in \cite{Wav2Vec2SER1, Wav2Vec2SER2}.

We actively utilized the NVIDIA Audio2Face \cite{audio2face} neural network for visualizing our results and collecting a dataset of human preferences.

We utilize Direct Preference Optimization (DPO) \cite{DPO}, a technique that advances the concept of Reinforcement Learning From Human Feedback (RLHF) \cite{FirstRLHFPaper, SummarizeRLHF}. This approach enables model training directly on human feedback without the necessity of training a separate reward model.

\begin{figure*}[th]
  \centering
  \includegraphics[width=\linewidth]{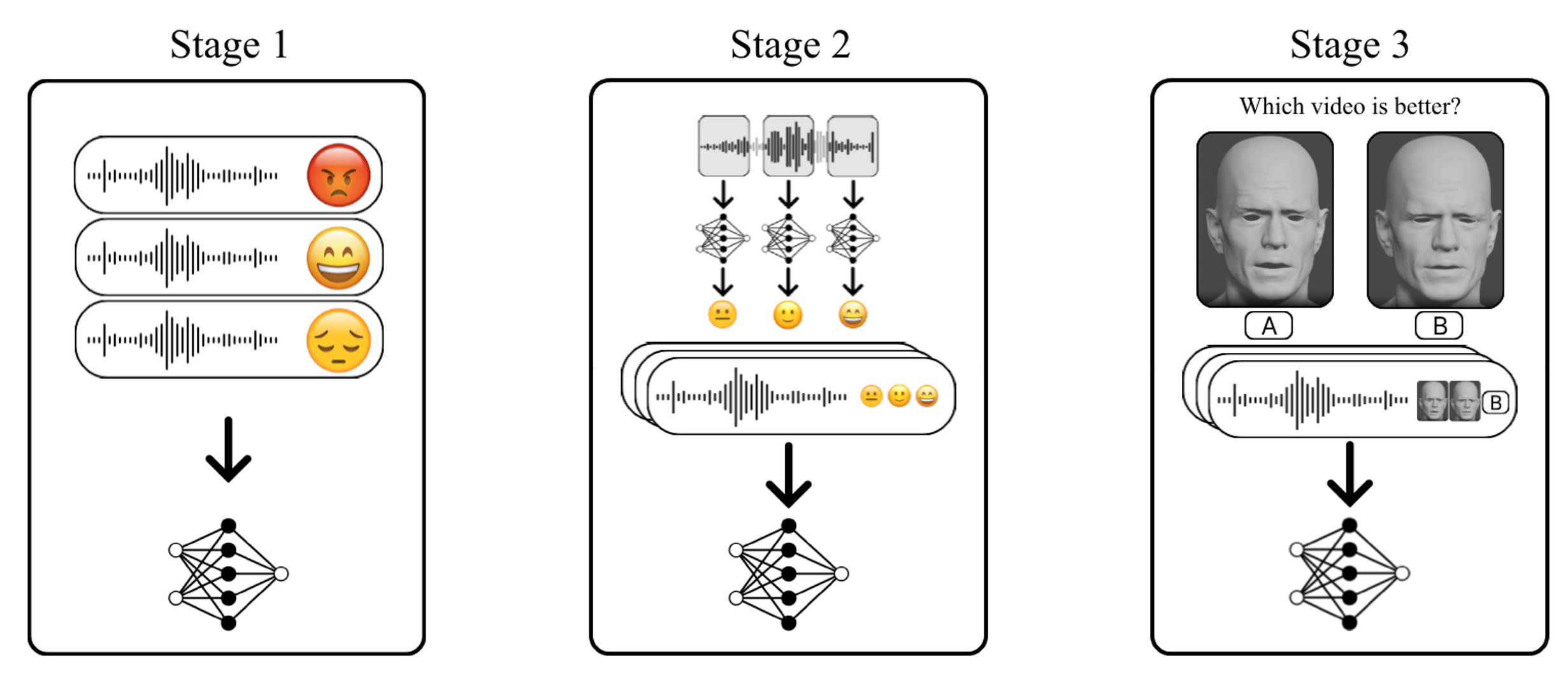}
  \caption{Three stages of the proposed method. At the first stage we train model A to predict a single emotion for the entire track. At the second stage we use the model A to generate sequential data using the sliding window approach and train model B on this data. At the final stage we further improve the model B by incorporating human feedback.}
  \label{fig:speech_production}
\end{figure*}

\section{Proposed method}

Our method is multi-staged. In the first stage, we train a neural network to predict a singular emotional representation for the entire audio track. In the second stage, we train a sequence-to-sequence model capable of predicting a sequence of emotions in one forward run, utilizing synthetic data generated by the model trained in the first stage. In the final third stage, we further fine-tune the model from the second stage using human feedback. These stages are illustrated in Figure 1.

Before delving into the detailed description of these stages, let us first consider our approach to formalizing the task of emotion recognition. In order to visualize the results of our experiments, we utilized a neural network to generate 3D facial animations based on voice audio and a sequence of emotions. This sequence is represented by numerical vectors ranging from 0 to 1, where each vector corresponds to a continuous description of an emotion at a specific time point. Each scalar within these vectors represents the degree of a particular emotion's activity, such as anger or joy. In this context, our interest lay in creating a probabilistic model for a mixture of emotions rather than probabilities of marginal categories. Unlike traditional classification tasks that presuppose a direct link between objects and distinct classes, we aim to associate an object with a categorical distribution reflecting a mixture of emotions. Formally, our goal is to develop a probabilistic model that estimates not the marginal probabilities of classes,
\begin{align*}
p (\text{emotion}\mid \text{audio}), \, \, \text{emotion} \in \{\text{anger, sadness, \dots, joy}\},
\end{align*}
but rather the joint probability of a mixture of emotions,
\begin{align*}
p(\text{anger} &= p_1, \text{sad} = p_2, \dots, \text{joy} = p_n \, | \, \text{audio}), \\
p_i &\in (0, 1), \quad \sum_{i=1}^N p_i = 1.
\end{align*}

We propose modeling emotion mixtures using the Dirichlet distribution, an N-dimensional continuous distribution parameterized by N positive values, $\alpha = \{ \alpha_i > 0\}_{i=1}^N$, with its domain being a simplex:
\begin{align}
\text{Dir} (x\mid \alpha)  &\propto \prod_{i=1}^N x_i^{\alpha_i - 1}, \quad x \in \mathcal{C}, \\ 
\mathcal{C} = \{ (x_1, x_2, \dots, x_N&) \in \mathbb{R}^N\, | \,x_i > 0, \,  \sum_{i=1}^N x_i = 1 \}.
\end{align}

\noindent This formulation implies that a sample from the Dirichlet distribution represents a categorical distribution. Thus, the mixtures of emotions we aim to model are effectively samples from a specific Dirichlet distribution. To define this distribution, we predict its parameters using a neural network $f$ with parameters $\theta$:
\begin{align}
\alpha &= f_\theta(\text{audio}) \\
\text{emotion} &\sim \text{Dir} (\text{emotion}\mid \alpha)
\end{align}

\noindent The model $f_\theta$ is trained by maximizing the likelihood across the training dataset $\{(\text{audio}_i, \text{emotion}_i)\}_{i=1}^K$:
\begin{align}
    \theta^* = \argmax_\theta \frac{1}{K} \sum_{i=1}^K \log \text{Dir} (\text{emotion}_i\mid f_\theta(\text{audio}_i)).
\end{align}

\noindent A model trained in this manner enables the estimation of probabilities for any emotion mixture. To infer a specific prediction, we use the expectation of the Dirichlet distribution:
\begin{align}
    \text{emotion}_i = \frac{\alpha_i}{\sum_{i=1}^N \alpha_i}.
\end{align}

\subsection{Modeling singular emotional representation}

In the first stage, we train the model to predict a singular mixture of emotions for the entire audio track. We solve this problem using classic supervised approach with objective (5). We utilized a dataset consisting of (audio, emotion) pairs, where the emotion is represented as a discrete one-hot encoded label. We used a set of six emotional categories: \textit{anger}, \textit{disgust}, \textit{fear}, \textit{joy}, \textit{neutral}, and \textit{sadness}. This selection was based on the availability of datasets in the public domain.

\subsection{Modeling sequence of emotions}
In dynamic speech emotion recognition, we aim to evolve from identifying a singular emotional vector for the entire audio track to predicting a sequence of emotions with corresponding time stamps. The obvious method for doing this transition is through the sliding window approach. This technique involves applying a model, trained at the 1st stage, to small and uniformly spaced segments of the audio, thereby constructing an emotional timeline. However, this method encounters two primary challenges: the limited contextual understanding due to the model processing only short waveform segments; and the inherent assumption of a fixed grid for the windows, restricting the time stamp placement of the emotions and requiring a search procedure to determine the optimal window size and stride.

To overcome these limitations, we suggest training a model capable of directly predicting an entire emotional sequence in one forward pass. This allows the model to observe the full context of the audio and precisely decide when to change the emotion. Due to the lack of datasets with time-specific emotion annotations, we created synthetic data through the sliding window method for this purpose. We applied the model trained on the first stage to generate such samples and then trained a new model to reproduce this behaviour.

\subsection{Incorporating human feedback}

At the final stage, we further fine-tune the model trained on the 2nd stage via learning from the human feedback paradigm. To do this, we generated multiple emotional sequences for each audio track, fed these into a facial animation neural network, and rendered videos showcasing digital humans exhibiting various emotional sequences. We then annotated these video pairs by selecting the preferred ones.

After the creation of the dataset, we fine-tuned the sequence-to-sequence model from the 2nd stage using Direct Preference Optimization (DPO) algorithm. The method utilizes the following loss for optimization:
\begin{align}
    \mathcal{L} = -\mathbb{E} \left[ \log \sigma \left( \beta \log \frac{\text{Dir}_\theta(y_w \, | \, x)}{\text{Dir}_{\text{ref}}(y_w \, | \, x)} -\beta \log \frac{\text{Dir}_\theta (y_l \, | \, x)}{\text{Dir}_{\text{ref}}(y_l \, | \, x)}\right)\right].
\end{align}

\noindent 
In this formula, the expectation is taken over triplets $(y_w, y_l, x)$ -- the preferred emotion, the dispreferred one, and the audio accordingly. $\text{Dir}_\theta(y_w \, | \, x)$ is the model to be trained w.r.t. the parameters $\theta$, while $\text{Dir}_{\text{ref}}(y \, | \, x)$ is the reference model explained in detail in the DPO paper. We used the frozen copy of the model from the 2nd stage as the reference model. The $\beta$ parameter controls how different the trainable model's behavior can be from the reference model. The loss (7) is calculated for each emotion in the sequence and then averaged. This formula makes it clear why the approach with the Dirichlet distribution was developed: if we only had a probabilistic model for marginal emotions, rather than their mixtures, we would have been unable to express the joint probability used in the formula.

\section{Experiments}

\subsection{Datasets and evaluation}
\textbf{Stage 1.} 
To increase data diversity and size, our training dataset was compiled from several publicly available sources and one private dataset, which we do not disclose due to commercial interests. The composition of the final dataset is detailed in Table \ref{tab:total_dataset}. It includes 18122 samples, corresponding to 15 hours of audio with an average duration of 3 seconds per audio clip. All recordings are in English and feature a gender balance of approximately 1:1.

\begin{table}[t]
  \caption{The details of the stage one training dataset.}
  \label{tab:total_dataset}
\begin{adjustbox}{width=\columnwidth,center}
\begin{tabular}{c|c|c|c|c|c|c|c}
                 & \textbf{anger} & \textbf{disgust} & \textbf{fear} & \textbf{joy} & \textbf{neutral} & \textbf{sadness} &      \\ \hline
\textbf{CREMA-D\cite{Cremad}} & 1271           & 1271             & 1271          & 1271         & 1087             & 1271             & 7442 \\ \hline
\textbf{RAVDESS\cite{ravdess}} & 192            & 192              & 192           & 192          & 96               & 192              & 1056 \\ \hline
\textbf{JL\cite{jl}}      & 240            & -                & -             & 480          & 240              & 240              & 1200 \\ \hline
\textbf{Private} & 1404           & 1404             & 1404          & 1404         & 1404             & 1404             & 8424 \\ \hline
                 & 3107           & 2867             & 2867          & 3347         & 2827             & 3107             & 18122    
\end{tabular}
\end{adjustbox}
\end{table}

To achieve a better assessment of model accuracy, we employed cross-dataset validation. We independently collected a test dataset consisting of 516 samples, with 86 samples for each emotion. The audio recordings were collected from 7 speakers, each of whom recorded half of the samples in English and the other half in their native language. We used standard classification accuracy as the metric for evaluation.

\textbf{Stage 2.}
To generate data with emotions that change over time we require an emotionally dynamic dataset. Standard datasets used in the first stage are unsuitable as they feature audio samples with a mostly static emotion. Instead, we utilized the VoxMovies \cite{Voxmovies1, Voxmovies2} collection, which comprises audio tracks of dialogues from movies and TV shows. We used a sliding window with a size of 1.4 sec and a step of 1 sec to predict the sequence of emotions for each audio track in this dataset, yielding a total of 676 sequence-to-sequence samples.

Assessing such a model poses a significant challenge. One approach to do that is to use the standard method of a hold-out sample and compare the predictions with pre-generated data. However, in this scenario, the metric values would only indicate how well the model replicates the sliding window method.

As previously mentioned, we had a pre-trained facial animation neural network capable of generating 3D digital avatars based on input audio and a sequence of emotions. Additionally, we had a small dataset of 3D facial animations and corresponding audio tracks. To create a ground truth dataset for evaluating our sequence-to-sequence model, we extracted emotional sequences from this dataset by optimizing the NVIDIA Audio2Face network (A2F) w.r.t. the input emotion. Namely, we froze the weights of A2F and found emotional sequences that minimized the Mean Squared Error (MSE) between the ground truth animation and the prediction by A2F. Formally, we solved the following optimization problem for each frame in the animation:
\begin{align}
    \text{emotion}^* = \argmin_{\text{emotion}} \text{MSE}(\text{A2F}(\text{audio}, \text{emotion}), \text{GT}).     
\end{align}
Using this approach, we extracted 40 emotional sequences directly from the ground-truth facial movements. While this data is insufficient for the training of a DSER model, it is enough for assessing its quality. We do not disclose this dataset due to commerical interest.

As a metric, Mean Absolute Error (MAE) was employed. The use of the Kullback-Leibler divergence was limited by the fact that the animating neural network utilized a different emotional space, notably where the zero emotional vector corresponded to a neutral emotion.

\textbf{Stage 3.}
To gather the human feedback, we utilized the manually filtered MELD \cite{MELD} dataset. This dataset represents speech recordings from a comedy show. Because of that it contains a lot of samples with background laughter, biasing the prediction towards the \textit{joy} emotion. We removed those samples, resulting in a total of 400 samples. For each audio track included, we generated 5 distinct sequences of emotions. Randomization involved applying the model from the 1st stage to the samples using the sliding window approach with varying window sizes and strides. These parameters were randomly selected from a range between \SI{0.25}{\sec} to \SI{1.25}{\sec}, under the condition that similar parameters could not repeat, to increase the diversity of generated samples. Subsequently, we rendered videos for each sequence of emotions. For the same audio, the annotator was shown 2 videos. The goal was to choose the one where the sequence of emotions was more preferable. In the result, we obtained 1500 samples of side-by-side comparisons of emotional sequences. We do not disclose this dataset due to commercial interests.

\begin{table}[t]
  \caption{Evaluation of the models predicting single emotion.}
  \label{tab:example}
  \centering
  \begin{tabular}{ l  l  l }
    \toprule
    \textbf{Loss} & \textbf{Backbone} & \textbf{Accuracy} \\
    \midrule
    Cross-Entropy         & Large~~~          & \textbf{0.84} \\
    Dirichlet Likelihood   & Large~~~            & 0.81 \\
    \midrule
    Cross-Entropy   & Base~~~            & \textbf{0.8} \\
    Dirichlet Likelihood   & Base~~~    & 0.79 \\
    \bottomrule
  \end{tabular}
\end{table}

\begin{table}[t]
  \caption{Evaluation of the sequential models.}
  \label{tab:example}
  \centering
  \begin{tabular}{ l  l  l }
    \toprule
    \textbf{Approach and loss} & \textbf{Backbone} & \textbf{MAE} \\
    \midrule
    Seq2Seq + DPO         & Large~~~          & \textbf{0.195} \\
    Seq2Seq + Dirichlet   & Large~~~            & 0.200 \\
    Seq2Seq + Cross-Entropy   & Large~~~            & 0.201 \\
    Sliding Window + Dirichlet            & Large~~~    & 0.206 \\
    Sliding Window + Cross-Entropy         & Large~~~           & 0.207 \\
    \midrule
    Seq2Seq + DPO         & Base~~~          & \textbf{0.203} \\
    Seq2Seq + Dirichlet   & Base~~~            & 0.206 \\
    Seq2Seq + Cross-Entropy   & Base~~~            & 0.208 \\
    Sliding Window + Dirichlet            & Base~~~    & 0.228 \\
    Sliding Window + Cross-Entropy         & Base~~~           & 0.231 \\
    \bottomrule
  \end{tabular}
\end{table}

\subsection{Model architecture.} In all our experiments we used Wav2Vec2.0. This model processes the audio waveform to generate a sequence of features. At the 1st stage we pool these features by averaging across time, creating a single vector for each audio clip. At the 2nd and 3rd stages we instead apply sliding window based smoothing with kernel=5 and stride=1. After that we apply a fully connected layer to reduce the feature dimensions to 6, matching the number of target emotions. These outputs are then transformed to positive values suitable for the Dirichlet distribution parameters via mapping $g(x) = x^2 + 10^{-6}$.

\subsection{Training details}
We experimented with two pretrained Wav2Vec2.0 checkpoints available through the Hugging Face \cite{Huggingface} platform: \textit{facebook/wav2vec2-base} (94M parameters) and \textit{facebook/wav2vec2-large-lv60} (315M parameters). Hereafter, we will refer to these models as Base and Large, respectively. 

We conducted approximately 200 runs to find the optimal hyperparameters for the 1st stage model, as this model was subsequently used for generating synthetic data. Validation was performed on a hold-out portion of our collected dataset. Overall, the model proved to be quite resilient to changes in hyperparameters. In the best checkpoint, we employed the following setup: 100 epochs, Adam optimizer with learning rate decaying from $5 \times 10^{-5}$ to $10^{-5}$, a batch size of 16, and a weight decay of $10^{-4}$. Additionally, we utilized augmentations such as audio shifting, cropping and noise injection. For subsequent stages, we used the same parameters. For all the experiments we resampled the input audio samples to \SI{16}{\kilo\hertz}.

\subsection{Results and discussion}

Tables 2 and 3 show the evaluation of the single emotion and sequence-to-sequence models, respectively. While the traditional cross-entropy loss shows slightly better results in terms of accuracy when predicting emotion for the entire audio, the Dirichlet method results in better metric values for sequential mixture modeling. The incorporation of the human feedback further improves the result, providing state-of-the art quality.

We examined the effect of the regularization parameter $\beta$ in the loss (7), noting that higher $\beta$ values align the model more closely with the reference model, as posited by DPO theory. This correlation was confirmed through empirical observation. Furthermore, excessively low $\beta$ values resulted in the model losing its pretrained characteristics, consequently diminishing performance metrics. Table 4 represents the influence of this parameter for the Large checkpoint. The value $\beta=0.5$ shows the optimal balance between learning from human feedback and utilizing the reference model's knowledge. The reported values represent the best metric scores achieved during training. 

Despite the human feedback-based method offering a simpler procedure for annotating and scaling datasets, in practice, it is sometimes challenging to compare the proposed emotional sequences for selection, which reduces the efficiency of labeling. This can be viewed as the method's limitation.

\begin{table}[t]
  \caption{The influence of the $\beta$ parameter.}
  \label{tab:beta_influence}
  \centering
  \begin{tabular}{lc}
    \toprule
    \textbf{$\mathbf{\beta}$} & \textbf{MAE} \\
    \midrule
    0.01 & 0.215 \\
    0.1 & 0.197 \\
    0.5 & \textbf{0.195} \\
    10.0 & 0.200 \\
    \bottomrule
  \end{tabular}
\end{table}

Our approach was trained and evaluated on a single node with an NVIDIA RTX 3090 GPU. The training of each stage takes 1-2 hours. Inference for the largest checkpoint, accelerated with NVIDIA TensorRT in FP16 format, takes about \SI{6}{\milli s} for \SI{1}{\sec} of input audio, enabling real-time usage of the model.

\section{Conclusion}

In this study, we introduced Dynamic Speech Emotion Recognition (DSER), a novel approach for over-time speech emotion recognition method suitable for 3D avatar animation. Our approach, leveraging synthetic data generation in conjunction with human feedback fine-tuning, aimed to overcome the challenges in traditional emotion recognition techniques. The results showcase the efficiency of our approach in accurately modeling sequences of emotional mixtures, and highlight the value of human feedback in enhancing model performance. We see further research in scaling datasets and more detailed control over emotion mixture modeling as promising directions.

\bibliographystyle{IEEEtran}
\bibliography{mybib}

\begin{thebibliography}{10}
\providecommand{\url}[1]{#1}
\csname url@samestyle\endcsname
\providecommand{\newblock}{\relax}
\providecommand{\bibinfo}[2]{#2}
\providecommand{\BIBentrySTDinterwordspacing}{\spaceskip=0pt\relax}
\providecommand{\BIBentryALTinterwordstretchfactor}{4}
\providecommand{\BIBentryALTinterwordspacing}{\spaceskip=\fontdimen2\font plus
\BIBentryALTinterwordstretchfactor\fontdimen3\font minus \fontdimen4\font\relax}
\providecommand{\BIBforeignlanguage}[2]{{%
\expandafter\ifx\csname l@#1\endcsname\relax
\typeout{** WARNING: IEEEtran.bst: No hyphenation pattern has been}%
\typeout{** loaded for the language `#1'. Using the pattern for}%
\typeout{** the default language instead.}%
\else
\language=\csname l@#1\endcsname
\fi
#2}}
\providecommand{\BIBdecl}{\relax}
\BIBdecl

\bibitem{audio2face}
\BIBentryALTinterwordspacing
T.~Karras, T.~Aila, S.~Laine, A.~Herva, and J.~Lehtinen, ``Audio-driven facial animation by joint end-to-end learning of pose and emotion,'' \emph{ACM Trans. Graph.}, vol.~36, no.~4, jul 2017. [Online]. Available: \url{https://doi.org/10.1145/3072959.3073658}
\BIBentrySTDinterwordspacing

\bibitem{DPO}
\BIBentryALTinterwordspacing
R.~Rafailov, A.~Sharma, E.~Mitchell, C.~D. Manning, S.~Ermon, and C.~Finn, ``Direct preference optimization: Your language model is secretly a reward model,'' in \emph{Thirty-seventh Conference on Neural Information Processing Systems}, 2023. [Online]. Available: \url{https://arxiv.org/abs/2305.18290}
\BIBentrySTDinterwordspacing

\bibitem{Wav2Vec2}
\BIBentryALTinterwordspacing
A.~Baevski, H.~Zhou, A.~rahman Mohamed, and M.~Auli, ``wav2vec 2.0: A framework for self-supervised learning of speech representations,'' \emph{ArXiv}, vol. abs/2006.11477, 2020. [Online]. Available: \url{https://api.semanticscholar.org/CorpusID:219966759}
\BIBentrySTDinterwordspacing

\bibitem{pepino21_interspeech}
L.~Pepino, P.~Riera, and L.~Ferrer, ``{Emotion Recognition from Speech Using wav2vec 2.0 Embeddings},'' in \emph{Proc. Interspeech 2021}, 2021, pp. 3400--3404.

\bibitem{Hubert}
\BIBentryALTinterwordspacing
W.-N. Hsu, B.~Bolte, Y.-H.~H. Tsai, K.~Lakhotia, R.~Salakhutdinov, and A.~Mohamed, ``Hubert: Self-supervised speech representation learning by masked prediction of hidden units,'' \emph{IEEE/ACM Trans. Audio, Speech and Lang. Proc.}, vol.~29, p. 3451–3460, oct 2021. [Online]. Available: \url{https://doi.org/10.1109/TASLP.2021.3122291}
\BIBentrySTDinterwordspacing

\bibitem{Wav2Vec2SER1}
\BIBentryALTinterwordspacing
L.~Pepino, P.~E. Riera, and L.~Ferrer, ``Emotion recognition from speech using wav2vec 2.0 embeddings,'' \emph{ArXiv}, vol. abs/2104.03502, 2021. [Online]. Available: \url{https://api.semanticscholar.org/CorpusID:233181984}
\BIBentrySTDinterwordspacing

\bibitem{Wav2Vec2SER2}
\BIBentryALTinterwordspacing
Y.~Wang, A.~Boumadane, and A.~Heba, ``A fine-tuned wav2vec 2.0/hubert benchmark for speech emotion recognition, speaker verification and spoken language understanding,'' \emph{ArXiv}, vol. abs/2111.02735, 2021. [Online]. Available: \url{https://api.semanticscholar.org/CorpusID:242757022}
\BIBentrySTDinterwordspacing

\bibitem{FirstRLHFPaper}
\BIBentryALTinterwordspacing
P.~F. Christiano, J.~Leike, T.~Brown, M.~Martic, S.~Legg, and D.~Amodei, ``Deep reinforcement learning from human preferences,'' in \emph{Advances in Neural Information Processing Systems}, I.~Guyon, U.~V. Luxburg, S.~Bengio, H.~Wallach, R.~Fergus, S.~Vishwanathan, and R.~Garnett, Eds., vol.~30.\hskip 1em plus 0.5em minus 0.4em\relax Curran Associates, Inc., 2017. [Online]. Available: \url{https://proceedings.neurips.cc/paper_files/paper/2017/file/d5e2c0adad503c91f91df240d0cd4e49-Paper.pdf}
\BIBentrySTDinterwordspacing

\bibitem{SummarizeRLHF}
N.~Stiennon, L.~Ouyang, J.~Wu, D.~M. Ziegler, R.~Lowe, C.~Voss, A.~Radford, D.~Amodei, and P.~Christiano, ``Learning to summarize from human feedback,'' in \emph{Proceedings of the 34th International Conference on Neural Information Processing Systems}, ser. NIPS'20.\hskip 1em plus 0.5em minus 0.4em\relax Red Hook, NY, USA: Curran Associates Inc., 2020.

\bibitem{Cremad}
H.~Cao, D.~G. Cooper, M.~K. Keutmann, R.~C. Gur, A.~Nenkova, and R.~Verma, ``Crema-d: Crowd-sourced emotional multimodal actors dataset,'' \emph{IEEE Transactions on Affective Computing}, vol.~5, no.~4, pp. 377--390, 2014.

\bibitem{ravdess}
\BIBentryALTinterwordspacing
S.~R. Livingstone and F.~A. Russo, ``The ryerson audio-visual database of emotional speech and song (ravdess): A dynamic, multimodal set of facial and vocal expressions in north american english,'' \emph{PLOS ONE}, vol.~13, no.~5, p. e0196391, May 2018. [Online]. Available: \url{http://dx.doi.org/10.1371/journal.pone.0196391}
\BIBentrySTDinterwordspacing

\bibitem{jl}
J.~James, L.~Tian, and C.~Watson, ``An open source emotional speech corpus for human robot interaction applications,'' 09 2018, pp. 2768--2772.

\bibitem{Voxmovies1}
A.~Brown, J.~Huh, A.~Nagrani, J.~S. Chung, and A.~Zisserman, ``Playing a part: Speaker verification at the movies,'' in \emph{International Conference on Acoustics, Speech, and Signal Processing (ICASSP), 2021}, 2020.

\bibitem{Voxmovies2}
A.~Nagrani, J.~S. Chung, W.~Xie, and A.~Zisserman, ``Voxceleb: Large-scale speaker verification in the wild,'' \emph{Computer Science and Language}, 2019.

\bibitem{MELD}
\BIBentryALTinterwordspacing
S.~Poria, D.~Hazarika, N.~Majumder, G.~Naik, E.~Cambria, and R.~Mihalcea, ``{MELD}: A multimodal multi-party dataset for emotion recognition in conversations,'' in \emph{Proceedings of the 57th Annual Meeting of the Association for Computational Linguistics}, A.~Korhonen, D.~Traum, and L.~M{\`a}rquez, Eds.\hskip 1em plus 0.5em minus 0.4em\relax Florence, Italy: Association for Computational Linguistics, Jul. 2019, pp. 527--536. [Online]. Available: \url{https://aclanthology.org/P19-1050}
\BIBentrySTDinterwordspacing

\bibitem{Huggingface}
\BIBentryALTinterwordspacing
T.~Wolf, L.~Debut, V.~Sanh, J.~Chaumond, C.~Delangue, A.~Moi, P.~Cistac, T.~Rault, R.~Louf, M.~Funtowicz, J.~Davison, S.~Shleifer, P.~von Platen, C.~Ma, Y.~Jernite, J.~Plu, C.~Xu, T.~Le~Scao, S.~Gugger, M.~Drame, Q.~Lhoest, and A.~Rush, ``Transformers: State-of-the-art natural language processing,'' in \emph{Proceedings of the 2020 Conference on Empirical Methods in Natural Language Processing: System Demonstrations}, Q.~Liu and D.~Schlangen, Eds.\hskip 1em plus 0.5em minus 0.4em\relax Online: Association for Computational Linguistics, Oct. 2020, pp. 38--45. [Online]. Available: \url{https://aclanthology.org/2020.emnlp-demos.6}
\BIBentrySTDinterwordspacing

\end{thebibliography}

\end{document}